\documentclass[
 reprint,
 amsmath,amssymb,
 aps,
 prb,
 floatfix,
]{revtex4-1}

\usepackage{graphicx}
\usepackage{dcolumn}
\usepackage{bm}
\usepackage{xcolor}
\usepackage{amsmath}

\newcommand{\ket}[1]{| #1 \rangle}
\newcommand{\bra}[1]{\langle #1 |}

\begin{document}

\preprint{APS/123-QED}

\title{A variational Monte Carlo approach for core excitations}

\author{Scott M. Garner$^{1,2}$}

\author{Eric Neuscamman$^{1,2,}$}%
\email{eneuscamman@berkeley.edu}

\affiliation{
${}^1$Department of Chemistry, University of California, Berkeley, CA, 94720, USA \\
${}^2$Chemical Sciences Division, Lawrence Berkeley National Laboratory, Berkeley, CA, 94720, USA
}

\date{\today}

\begin{abstract}
We present a systematically-improvable approach to core excitations
in variational Monte Carlo.
Building on recent work in excited-state-specific Monte Carlo,
we show how a straightforward protocol, starting from a quantum chemistry
guess, is able to capture core state's strong orbital relaxations,
maintain accuracy in the near-nuclear region during these relaxations,
and explicitly balance accuracy between ground and core excited states.
In water, ammonia, and methane, which serve as prototypical representatives
for oxygen, nitrogen, and carbon core states, respectively,
this approach delivers accuracies on par with
the best available theoretical methods even when using relatively small
wave function expansions.
\end{abstract}

\maketitle

\section{Introduction}
\label{sec::intro}

Quantum Monte Carlo (QMC) approaches
\cite{SorellaQMCBook,FoulkesMitasNeedsRajagopal,kolorenvc2011,austin2012}
have long been used to provide
highly-accurate theoretical reference data from which other,
less expensive methods can benefit.
Since the pioneering work of Ceperley and Alder \cite{CeperleyAlder}
that led to the development of the local density approximation,
there have been many cases where QMC methods have provided benchmarks
for and insights into other theories.
Examples include the phase stability of high-pressure hydrogen,
\cite{mcmahon2012}
the low-temperature properties of the Hubbard model, \cite{hubbardPRX2015}
the stability of a covalently-bound O$_4$ molecule, \cite{caffarel2007_O4}
and the optical gap of ZnO. \cite{zhao2019opticalgaps}
While the computational costs associated with taking large
random samples can be intimidating, the various types of QMC methodology
offer highly accurate and systematic predictions when
paired with parallel computing.

In particular, QMC allows increasingly large expansions of Slater
determinants to be coupled with correlation factors,
\cite{clark2011,morales2012,filippi2016derivatives,Sergio:2019,dash2019}
which allows both strong and weak correlation effects
as well as basis set effects to be addressed simultaneously,
especially in ground state contexts where projector Monte Carlo methods
can be used to polish off the finer details. \cite{scemama2016}
Crucially, the ability of the correlation factors (also called Jastrow
factors) to deal with electron cusps and other weak-correlation effects
means that accuracy is reached with far smaller determinant expansions
than are necessary in quantum chemistry.
In many cases, as in molecular O$_4$, even relatively short expansions
are sufficient to allow a clear balance to be established between
the accuracy of different states \cite{Robinson:2017,dash2019}
or the same state at different molecular geometries. \cite{caffarel2007_O4}
This balance enables by-design error cancellation and the delivery
of high-accuracy energy differences, even in cases like
O$_4$ where the wave functions at different geometries differ
substantially in the character of their electron correlation.
\cite{caffarel2007_O4}

One must bear in mind, however, that most success stories in QMC have
been achieved by carefully removing any
core electrons from the QMC simulation via pseudopotentials.
Unlike quantum chemistry, where freezing core electrons or using
pseudopotentials lowers the cost of correlation methods by a modest factor,
the efficiency gain in QMC can be quite dramatic due to both the need to
take smaller sampling steps when the kinetic energy scale is higher and
the higher energy variance (and thus statistical uncertainty) that
comes with the higher energy scale of core electrons.
For example, the computational cost of all-electron diffusion Monte Carlo has
been estimated to grow as roughly the sixth power of the nuclear charge.
\cite{ceperley1986,hammond1987}
Given the recent advances in excited-state QMC methods
\cite{filippi2009SAVMC,blunt2015fciqmcexcited,Chris:2016:omega,
Robinson:2017,Jacki:2017:sc,blunt2018,dash2019}
and the electronic structure community's continuing efforts
\cite{Gill:DSCF:Uncontract,
Coriani:2015,
CVSADC3,
Krylov:2019,
OOST:2018,
Xiaosong:2019,
Prendergast:2019,
oosterbaan2019,
peng2019,
Krylov:2020,
OOST:2020,
Dip:2020:roks_core,
Dip:2020:radical_core}
to develop affordable theoretical methods for core excitations,
it is interesting to ask whether the challenges that QMC faces for
core electrons can be overcome so that for relatively light elements
it can act as a reliable benchmark for more affordable theories,
just as it has in other areas.

Interest in core excitations and X-Ray absorption
spectroscopy (XAS) comes in large part
from the chemical analysis then can offer through
element, chemical environment, and spatial specificity.
\cite{XRay:WaterCoordination,XRay:DNA1,XRay:DNA2,Leone:RingOpening,Baker:XUV}
Recent improvements in X-Ray light sources' temporal, spatial, and spectral
resolutions \cite{FEL:Book} have enhanced these advantages and
opened the possibility for novel experiments
in nuclear and electronic dynamics. \cite{Young:2018,Leone:2018}
With these increasing capabilities comes an increasing need for reliable
theoretical methods that can unambiguously assigning experimental
features of ever-more-exotic core spectra.
Amidst the recent flurry of theoretical development in this area
\cite{Gill:DSCF:Uncontract,
Coriani:2015,
CVSADC3,
Krylov:2019,
OOST:2018,
Xiaosong:2019,
Prendergast:2019,
oosterbaan2019,
peng2019,
Krylov:2020,
OOST:2020,
Dip:2020:roks_core,
Dip:2020:radical_core}
we are eager to explore what QMC has to offer, beginning with its
traditional role as a theoretical benchmark.
In this study, we will address various challenges standing in the way
of high-accuracy core-state QMC through a variational Monte Carlo (VMC)
framework and provide proof-of-principle results in water, ammonia,
and methane (representing the O, N, and C K-edges)
demonstrating the promise of the approach.








\section{Theory}
\label{sec:theory}

\subsection{Overview}
\label{sec::overview}

In pursuing a VMC-based, systematically improvable approach to core excitation
energies and their corresponding wave functions,
there are a variety of issues that must be considered and addressed.
First, as evidenced by the successes of theories that have it
(e.g.\ NOCIS, \cite{OOST:2018}
$\Delta$SCF, \cite{Gill:DSCF:Uncontract}
and ROKS \cite{Dip:2020:roks_core})
and the failures of theories that don't
(e.g.\ configuration interaction singles\cite{OOST:2018}),
post-excitation orbital relaxations are essential when aiming for
high accuracy in core excitation energies.
Indeed, the accuracy of Hartree-Fock-based $\Delta$-SCF predictions of
core electron binding energies, \cite{Gill:DSCF:Uncontract}
which give results within an eV of experiment
despite neglecting 10 or more eV of correlation energy,
strongly argue that one should, as in NOCIS, \cite{OOST:2018}
not even worry about correlation effects
until orbital relaxations have been sorted out.
There are many ways one could address orbital relaxation in a VMC approach,
such as incorporating them in the quantum chemistry starting point before
even getting to VMC.
However, to ensure we are providing a stringent
test of VMC's ability to treat core states, we will in the present study
start from ground state Hartree-Fock orbitals and rely on recent advances
in VMC orbital optimization \cite{filippi2016derivatives}
to capture orbital relaxation effects.

A related challenge is how to optimize ground and core excited states
within the same VMC framework.
Some existing methods that work well for valence states,
such as state-averaged VMC, \cite{filippi2009SAVMC}
will not be appropriate for core excitations due to the strong
orbital relaxations involved.
Variance minimization \cite{Umrigar:1988} is appropriate in principle,
but relies on the initial guess being close to the desired state
as all Hamiltonian eigenstates have a the same (zero) variance.
In practice, the use of an approximate ansatz can make some states
stronger attractors than others when minimizing the variance
\cite{Filippi:2020}
as the variance minima around different states
will now have different depths.
If the ansatz is too poor, some minima may disappear altogether.
To guard against these issues, which we did see during
some early and insufficiently-cautious attempts at core state optimization,
we employ a variance minimization approach that is state-specific at
every individual optimization step \cite{Jacki:2017:sc} and enforce
stronger safeguards on it than in previous work (see Section \ref{sec::vmc}).

A third and obvious challenge is that pseudopotentials are not appropriate
for our purposes, at least not on the atom promoting a core electron.
Although all-electron calculations are certainly not unheard of in ground
state VMC (see e.g.\ work by Toulouse and Umrigar \cite{toulouse2007}),
we must take care here to treat the nuclear cusp and, more importantly,
\cite{Ma:Hats}
the wave function in its immediate vicinity carefully, as the local
electric field and wave function changes significantly after the excitation.
Unlike in ground state work, we should not expect errors in the treatment
of this region to cancel when we take energy differences, and so treating
this region accurately is a crucial.
As discussed in Section \ref{sec::basisandcusp}, we address this region through
a combination of modifying the basis set and adding an extra short-range
electron-nuclear Jastrow factor in addition to one with a more standard
range.


Fourth and finally, QMC methods that are treating substantially different
states must take care that their wave functions are of similar accuracy
so that energy differences will not be biased.
This issue is in some ways more severe than in basis-set-bound quantum
chemistry, where the idea of a model chemistry has real value in that the
basis set puts a hard limit on how much correlation energy can be recovered.
So long as the unrecoverable correlation energy is similar in all states
considered, cancellation of error is to some degree built in.
In QMC, by contrast, sophisticated Jastrow factors and especially
(although we do not employ it here) projector Monte Carlo methods make it
possible, in principle, to reach 100\% correlation recovery.
This promise can be a double-edged sword, however, when states of very
different character are being considered.
If QMC is much better at correlation recovery in one case than the other,
then energy differences can be biased to an extend that would be hard to
manage within the guard rails of a finite basis set.
As we are considering core excitations, in which substantial changes are
being made among the electrons with the largest energy scales,
caution in this regard is called for.
This is especially true given that we are aiming for high accuracy,
and thus using aggressive tools like three-body Jastrow factors
and multi-Slater expansions to aid in correlation recovery.
If our efforts are more effective in one state than the other,
and let's face it, most of these tools have been designed and used
in ground states, then we may inadvertently introduce a bias in our
quest for correlation recovery.
As discussed in Section \ref{sec::vm}, we will address these
concerns by employing the energy variance as a measure of wave function
accuracy to help ensure that our energy differences are balanced.


To summarize, the approach to core states that we pursue here
is designed to deal with large orbital relaxations,
convergence to the correct state, excitation-induced
changes in the near-nuclear region, and balancing accuracy.
The resulting protocol, which we will describe in detail
in the following sections, can be organized into four stages.
\begin{enumerate}
    \item Choice of basis set and 1-body Jastrow
    \item Guess preparation via quantum chemistry
    \item State-specific variational Monte Carlo
    \item Variance matching
\end{enumerate}


\subsection{Basis set and cusp considerations}
\label{sec::basisandcusp}

\begin{figure}
    \centering
    \includegraphics[width=0.48\textwidth]{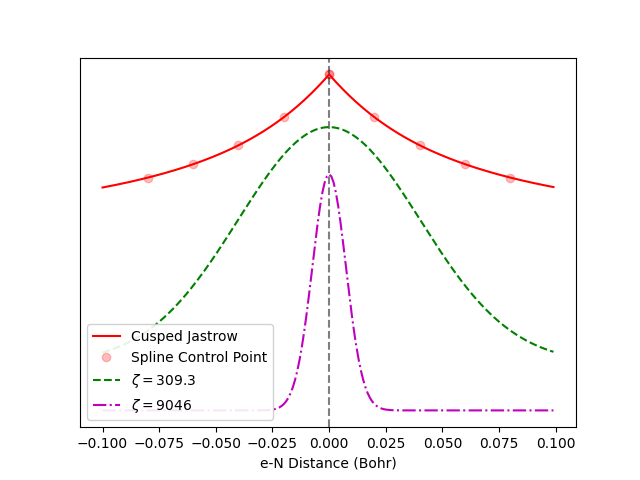}
    \caption{
    A snippet of the short range cusped jastrow factor on about the Nitrogen nucleus in Ammonia.  The spacing of the spline control points is indicated on the Jastrow.  Note this Jastrow is radially symmetric and only defined for $\vec{r}_{e,N}>0$, but we have plotted both positivie and negative distances to show the cusped behavior more clearly.  Also plotted are two Gaussians with $\zeta = 309.3$ and $\zeta = 9046$ which are the tightest function allowed and a function removed from the uncontracted basis, respectively.  The Gaussians are not normalized and have been translated vertically.
    \label{fig:1jc}
    }
\end{figure}

As emphasized by Gill et al \cite{Gill:DSCF:Uncontract}
and more recently by Krylov et al, \cite{Krylov:2020}
uncontracting a standard basis set can be helpful when
modeling core excitations.
If anything, uncontraction can be even more helpful
in the VMC context, as it allows one to eliminate variable
redundancies and corresponding optimization difficulties
in the near-nucleus region.
As one-electron Jastrow factors can easily encode
the analytically-known \cite{Kato:Cusps} electron-nuclear
cusp at the nuclei as well as the shape of one-electron
functions in its immediate vicinity, large Gaussian
contraction schemes that work to shape this region
ever closer in to the nuclei are not only unnecessary,
but in fact problematic.
To see why, consider Figure \ref{fig:1jc}, where we
show two relatively tight Gaussian primitives alongside
a one-electron Jastrow factor build from cubic B-splines.
\cite{QMCPACK:2018}
Consider for now the ground state, where we can
assume that a large contraction of Gaussian primitives will
have the orbital shape more or less correct except for the
in the region extremely close to the nucleus.
In this case, the Jastrow factor will need to switch from
having nontrivial structure in the tiny region around
the nucleus where the Gaussian-type orbital needs correcting
to being essentially flat in the rest of space,
where the orbital is already correct.
The better the quality of the Gaussian-type orbital,
the shorter the distance over which this switch must occur,
and thus the tighter the spline grid will need to be
near the core.
Worse, to actually optimize the Jastrow in this tiny region,
extremely large random samples will be required, as seeing
an electron land in the region where the Jastrow matters
will be a rare event.

These issues have been recognized before,
and in cases like the ground state where the orbital
shape is essentially correct outside the tiny cusp region,
numerical methods have been developed to convert a
Gasssian-type orbital into something more Slater like
before many-electron QMC even gets started
by slightly modifying the orbital shape at the very center.
\cite{Jastrow:Hats:Old,Ma:Hats}
In our approach, however, the core excited state will
not have optimal orbital shapes until late in the VMC
optimization, and so this type of rigid before-QMC correction
is less appropriate.
Instead, we follow the uncontracted basis approach of Nakano
\textit{et al} and, as they recommend, remove the tightest Gaussian
primitives ($\zeta\ge 300$) from the basis. \cite{NaDimerCusps} 
To deal with near-nucleus orbital shaping that the basis can no longer deliver,
we employ a short range electron-nuclear Jastrow that effects only the
first 0.5 Bohr about the nucleus and contains 25 spline
points (15 for hydrogen atoms).
Note that this is a far less dense spacing than is required to make
the switch discussed above in the case of a GTO with very tight primitives,
and in practice we find it does not cause issues in our optimization.
Of course, the appropriate range and spline-point density of this Jastrow will
depend on how aggressive one is with the primitive-removal cutoff.
Note that we also include a second electron-nuclear Jastrow
that is cusp-free and longer ranged, with 25 spline points over 7.5 Bohr
for heavy atoms and 15 spline points over 7.5 Bohr for hydrogen atoms.


Starting from an aug-cc-pVDZ basis on non-hydrogen atoms,
we uncontract the basis functions and remove tight Gaussian functions.
For hydrogen, we use a standard contracted cc-pVDZ basis, as
moving to high-quality hydrogen basis sets is known to have little
impact on accuracy in this context. \cite{Krylov:2020}
While the resulting basis is small by core state quantum
chemistry standards, it is effective here for three reasons.
First, the use of highly-flexible short range electron-nuclear Jastrows
obviates the need for near-core flexibility in the basis.
Second, in addition to 1-body electron-nuclear Jastrows, we also
include 2-body electron-electron Jastrows \cite{QMCPACK:2018}
and the 3-body Jastrow of Needs
\textit{et al}. \cite{Drummond:3JFunctionalForm}
These additional correlation factors provide both
exact electron-electron cusps and encode some weak correlation,
reducing the need for the basis set to facilitate these through
the configuration-interaction expansion.
Third, the by-design error cancellation that variance matching
facilitates should help address any accuracy bias that a too-small
basis inflicts against core excited states.


\begin{figure}[t]
    \centering
    \includegraphics[width=0.30\textwidth]{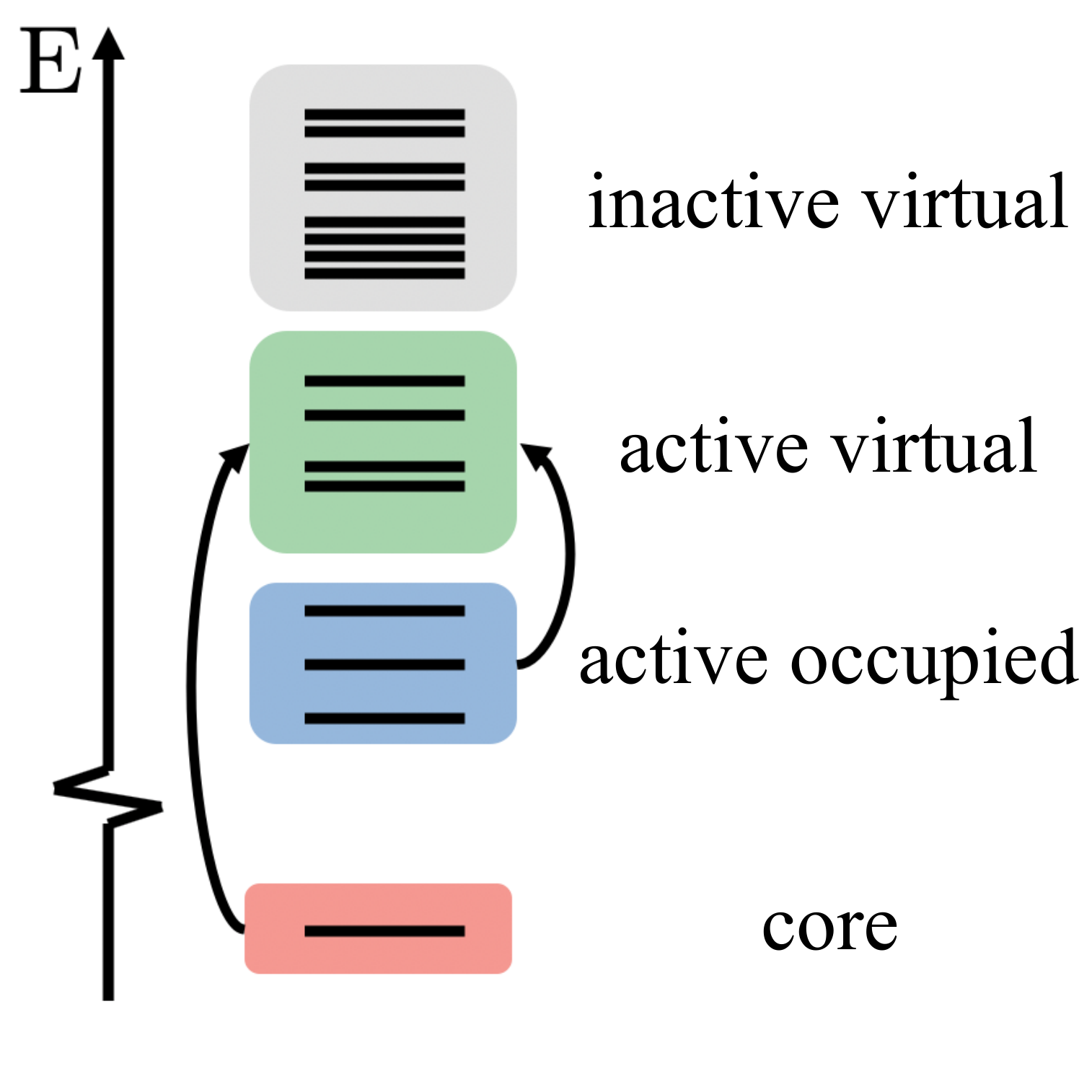}
    \caption{
    Orbital partitioning scheme for ORMAS.
    The core is held doubly-occupied in the ground state
    and singly-occupied in the excited states.
    Up to three additional transitions are allowed from
    active occupied to active virtual orbitals.
    \label{fig:RAS}
    }
\end{figure}

\subsection{Quantum Chemistry}
\label{sec::qc}


Our basis set chosen, we prepare a multi-determinant expansion
using the Occupation Restricted Multiple Active Spaces (ORMAS)
configuration interaction (CI) method \cite{Panin:RASCI,Ivanic:ORMAS} 
as implemented in GAMESS. \cite{GAMESS:1,GAMESS:2}
As shown in Figure \ref{fig:RAS}, we partition the ground state
restricted Hartree-Fock (RHF) orbitals into four categories.
For our ground state wave function, we enforce double occupation
of the core orbital and allow up to three excitations within
the active orbitals.
For the excited state, we require that one core electron
be promoted into the active virtual orbitals, and allow up to
three additional excitations from the active occupied
to the active virtual orbitals.
In this study, the active occupieds are defined as the
non-core occupied orbitals, while the active virtuals consist of the
remaining valence-shell orbitals
plus the next 9 virtual orbitals,
which in the systems studied here is a simple way to include
the 3s, 3p, and 3d of the heavy atom.
Note that, while the inactive virtual orbitals do not participate at the
quantum chemistry stage, they are carried over into VMC so that they can
participate in the orbital optimization.
In future, it will likely be preferrable to perform the excited state
ORMAS CI calculation in a core-relaxed basis set, which could be generated
via STEX, \cite{STEX:1997}
NOCIS, \cite{OOST:2018,OOST:2020}
ROKS, \cite{Dip:2020:roks_core}
or excited state mean field theory. \cite{hardikar2020}
That said, we start from RHF orbitals here in order to create
a more stringent test of VMC's ability to optimize core states.

One way to look at our ORMAS approach is as a particularly aggressive choice of
core-valence separation (CVS) scheme \cite{Cederbaum:CVS,Coriani:2015} 
that makes both the ground and excited state CI calculations variationally stable.
Crucially, and this aspect carries over to VMC as the correlation factors
are not so flexible as to refill the core,
the lack of filled-core configurations within the excited state
ansatz disables Auger coupling to the valence continuum.
Less happily, the rigid core occupations in both states
prevents the CI expansion from capturing core-core and non-Auger core-valence
correlation effects. \cite{Krylov:2019}
This would be more concerning if we were stopping at the quantum chemistry
stage, but within VMC these effects will be at least partially captured
by the two- and three-body Jastrow factors.



\subsection{Variational Monte Carlo}
\label{sec::vmc}

With a multi-determinant expansion in hand, we add the
Jastrow factors to complete the ansatz and proceed with our
state-specific VMC optimization, all of which is handled
by a development version of the QMCPACK software package.
\cite{QMCPACK:2018,QMCPACK:2020}
The full ansatz can be written as
a product of the Jastrow factor and a truncation
(see Section \ref{sec::vm} for truncation details)
of the quantum chemistry determinant expansion.
\begin{equation}\label{eqn:msj}
    \ket{\Psi}=e^{-J(\vec{r})}\sum_Ic_I\ket{\Psi_I(\vec{r})}
\end{equation}
Here each $\ket{\Psi_I(\vec{r})}$ is a Slater determinant with
associated CI coefficient $c_I$, and $e^{-J(\vec{r})}$ is the combined
one-, two-, and three-body Jastrow correlation factor.
\begin{subequations}
\begin{equation}
    J(\vec{r}) = J_1(\vec{r}) + J_2(\vec{r}) + J_3(\vec{r})
\end{equation}
\begin{equation}
    J_1(\vec{r})=\sum_{I,i,\sigma}
    U_I(|\vec{r}_{i,\sigma}-\vec{R}_{I}|) + V_I(|\vec{r}_{i,\sigma}-\vec{R}_{I}|)
\end{equation}
\begin{equation}
    J_2(\vec{r})=\sum_{i,j,\sigma,\tau}
                 W_{\sigma \tau}(|\vec{r}_{i,\sigma}-\vec{r}_{j,\tau}|)
\end{equation}
\end{subequations}
Here $r_{i,\sigma}$ is the position of the $i$th
spin-$\sigma$ electron, $R_I$ is the position of nuclei $I$,
and the $U$, $V$, and $W$ functions are cardinal cubic B-splines
with optimizable spline coefficients \cite{QMCPACK:2018}
for the short- and long-range one-body Jastrow and the two-body
Jastrow, respectively.
Note that, for $W$, there are two sets of spline coefficients, one for
like-spin electrons and one for opposite-spin electrons.
Kato's cusp conditions \cite{Kato:Cusps} are enforced explicitly
by the $U$ and $W$ functions.
Finally, for $J_3$, we use the functional form of Needs et al.
\cite{Drummond:3JFunctionalForm}


We optimize our ansatz using a modified
\cite{Chris:2016:omega,Jacki:2017:sc,QMCPACK:2018}
linear method \cite{Umrigar:2007,toulouse2007,Toulouse:2008}
implementation in a staged minimization of the objective function
\begin{equation}\label{TF}
    \Omega =\frac{\bra{\Psi} (\omega - H) \ket{\Psi}}{\bra{\Psi} (\omega - H)^2 \ket{\Psi}}
\end{equation}
which, for an exact ansatz, targets the first Hamiltonian
eigenstate with energy above the value $\omega$.
\cite{Chris:2016:omega,Jacki:2017:sc} 
If the approximate ansatz is accurate enough
that each Hamiltonian eigenstate in the relevant region
of variable space has its own separate variance minimum
(as would be the case for an exact ansatz),
and if the uncertainty in the Monte Carlo integration
with which we estimate the objective function and its derivatives
is low enough, then an optimization in which the initial guess
starts within the basin of the desired variance minimum and in which
$\omega$ is chosen and adjusted appropriately \cite{Jacki:2017:sc}
will converge to the variance minimum corresponding to the
desired state.
In practice, of course, some of these conditions may not be met,
in which case the optimization is at risk of converging
to a variance minimum that corresponds to a different
Hamiltonian eigenstate than the one desired. \cite{Filippi:2020}

In the present study, where the excited state ansatz starts with
unrelaxed orbitals and with Jastrow factors that have no structure
aside from the explicitly-enforced cusp conditions, it is hard
to argue that the initial guess and statistical precision are
good enough to avoid trouble.
Indeed, if we immediately start optimizing all orbital, CI,
and Jastrow parameters simultaneously with a moderate
sample size, the optimization typically fails to reach the
desired variance minimum.
Instead, we start by optimizing the one- and two-body Jastrow factors,
CI coefficients, and orbital rotation parameters for just
a few linear method steps, over which we observe a large
reduction in the energy variance
(from order 100 a.u.\ to order 1 a.u.).
It is tempting to be even more cautious by initially optimizing only
the  one- and two-body Jastrows,
but we are relying upon interplay between
the one-body Jastrow and the orbital coefficients
to get the near-nuclear region right.
Thus, we perform the first few iterations with everything except the three-body Jastrow enabled.
After this brief initial optimization, our statistical precision
is much improved and it is safer to begin
optimizing the more difficult three-body terms.
At this point, optimization continues in stages in which
different subsets of parameters are allowed to vary.
For example, we might optimize the orbitals with the other
parameters held fixed, then the 3-body Jastrow and the CI
coefficients with the orbitals held fixed, and so on.
For more difficult optimizations, particularly those with larger
determinant expansions, we find that starting from the
pre-optimized parameters of a smaller determinant expansion
for the same state is helpful.
In the final stage of the optimization, all parameters are
varied together.
The last iterations of this stage are then used for
variance matching, as described in the next section.

As has long been practiced for the ground state linear method,
our implementation uses an independent
sample to evaluate whether the objective function is actually
lowered by any of the three different proposed update steps
(one for each setting of the stabilizing shifts\cite{QMCPACK:2018})
and rejects any update that raises the objective function.
To help avoid large fluctuations in the variance and objective
function, we employ a clipping scheme, \cite{Runge:1993}
which serves much the same purpose as a more statistically
rigorous modified guiding function \cite{Robinson:2017} 
while being far easier to implement.
Note that $\omega$ is treated as a constant during a given linear
method iteration --- it is only ever updated in between linear
method evaluations --- and so the objective function that is making
the decision to accept or reject a parameter update is
state-specific, which is a key difference compared to
straightforward variance minimization. \cite{Umrigar:1988}
Even with our carefully staged optimization, we find that,
if we do not employ this rigorous check-and-reject step,
some of our optimizations can wander off to other states,
adding additional support to the concerns raised by
Filippi and coworkers \cite{Filippi:2020}
and highlighting the value of developing more robust
state-specific approaches.
We note that adapting a hybrid optimization scheme \cite{Leon:BLM}
to work with excited state objective functions
has proven effective in this regard, \cite{Leon:2020}
but that work is ongoing and is not the focus
of the present study.

\begin{figure*}[t]
    \centering
    \includegraphics[width=0.90\textwidth]{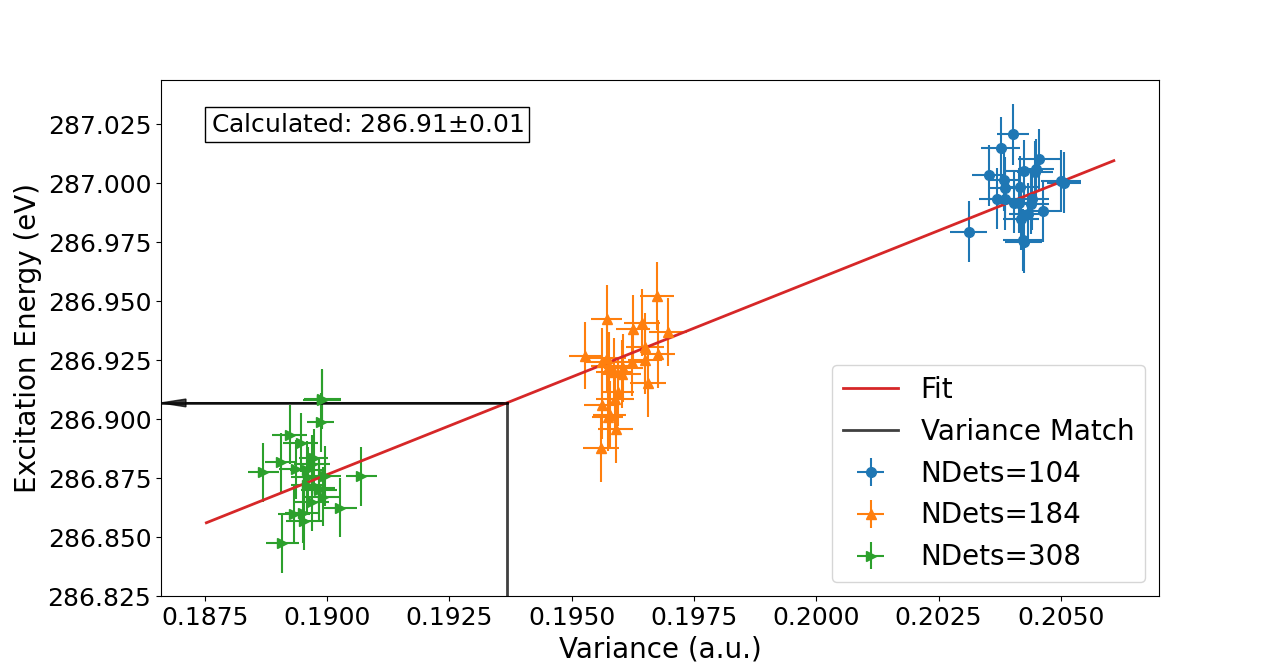}
    \caption{
    \label{fig:vm_methane}
    A demonstration of our variance matching approach in the
    case of methane's lowest core excited state.
    Each point gives the energy (relative to the ground state)
    and variance of the ansatz at one of the last 24 iterations of
    a wave function optimization, i.e.\ at the end of the final stage
    in which all variational parameters are optimized together.
    The three groups of points represent three different ansatzes,
    in which we retained 104, 184, and 308 determinants from
    the ORMAS wave function so as to straddle the ground state
    variance and allow for interpolation.
    A simple linear regression on all 72 points permits a
    straightforward interpolation of what the excited state's
    energy would be if the excited state exactly matched the
    ground state variance, as shown by the arrow.
    }
\end{figure*}

\subsection{Variance Matching}
\label{sec::vm}

Rather than assuming that our ground and excited state wave
functions are of equal quality and thus likely to
produce excitation energies that benefit from error cancellation,
we employ variance matching \cite{Robinson:2017}
to help ensure balance and accurate energy differences.
The variance
\begin{align}
    \sigma^2 = \left < ( \hat{H} - E )^2 \right >
\end{align}
is non-negative for any wave function and is only zero
for an exact Hamiltonian eigenstate, making it a strong
measure of an approximate wave function's quality.
Of course, lower variances are better, but when taking
energy differences, it seems equally important that
the states in question be balanced, and so there is an
argument for intentionally limiting one state's flexibility
so as to prevent it from being much better treated than
the other and thus biasing energy differences.
It is worth noting, though, that in using variance matching
to encourage error
cancellation, one is tacitly assuming that the states'
energy errors have the same sign.
Although there is no rigorous guarantee of this,
a qualitatively-correct ansatz for a low-lying state will
tend to be in error by containing many small contributions
from high-energy eigenstates, and so typically errors
relative to a chemically relevant state's true energy
will be positive.
In approaches that selectively take the most important
low-lying determinants to construct their ansatz, this
happy situation is even more likely to be true, as the
ansatz is by design missing contributions from only
higher-lying determinants.

Now, core states are not low-lying in the energy spectrum,
but the same basic logic should apply to them.
The only determinants that are lower in energy than those
used in our excited state determinant expansion are those
that have a filled core, and although there is some
Auger-like correlation energy associated with them,
it is typically less than 0.1 eV. \cite{Coriani:2015}
This is  much smaller than the correlation energy we are
missing due to an incomplete capture of correlation
effects associated with higher-lying determinants.
Furthermore, these Auger correlations are at least partially
captured through our two- and three-body Jastrow factors.
Thus, although there are some correlation terms that may work
to push our core state towards erroring low, these are expected
to be vastly outweighed by correlation terms that push us
towards erroring high.
Indeed, Figure \ref{fig:vm_methane} clearly shows that
our excited state energy decreases as we improve the
ansatz and decrease its variance.

\begin{table*}[t]
\caption{Transition energies (eV) for core excited states in
         methane, ammonia, and water.
         Experimental transition energies have been adjusted
         to remove relativistic effects \cite{RelativisticCorrections}
         and in one case a hot vibrational quanta.
         \cite{10e:EXP}
         Theoretical results (for both ground-to-excited
         and excited-to-excited transition energies)
         are reported as errors relative to the adjusted
         experimental numbers.
         For the VMC calculations, molecular geometries
         were taken from CCCBDB.nist.gov.
         Values for fc-CVS-EOM-CCSD
         are for the largest-basis
         calculations in Figures 2 and 3
         of Vidal et al. \cite{Krylov:2019}
         ROKS/SCAN values for NH$_3$ are
         taken (for precision's sake) from Figure 3 of
         Hait and Head-Gordon, \cite{Dip:2020:roks_core}
         while other ROKS/SCAN values
         are taken from the corresponding supporting information.
         VMC statistical uncertainties are less than 0.01 eV.
         \label{tab::energies}
        }
\begin{tabular}{l l r@{.}l r r@{.}l c c c}
\hline \hline
  \multicolumn{2}{c}{State\rule{0pt}{3.7mm}} &
  \multicolumn{2}{c}{\hphantom{.} Experiment$^a$ \hphantom{A}} &
  \multicolumn{1}{c}{VMC} &
  \multicolumn{2}{c}{ROKS/SCAN \cite{Dip:2020:roks_core}} &
  fc-CVS-EOM-CCSD\cite{Krylov:2019} &
  CVS-LR-CCSD\cite{CVS:Lanczos} &
  NOCIS\cite{OOST:2018} \\[0.5mm]
  \hline 
\multicolumn{10}{l}{ \textit{transition energies from the ground state}:\rule{0pt}{4.4mm} } \\[2mm]
CH$_4$ & 1s $\rightarrow$ 3$a_1$/3s & \hspace{4mm} 286&60$^{b,c}$ &  0.31 & \hspace{7mm} -0&2  &    -  &   -   &     - \\[0.4mm]
       & 1s $\rightarrow$ 2$t_2$/3p & \hspace{4mm} 287&90$^{c}$   &  0.21 & \hspace{7mm}  0&0  &    -  &   -   & -0.63 \\[0.4mm]
NH$_3$ & 1s $\rightarrow$ 4$a_1$/3s & \hspace{4mm} 400&45$^{d}$   &  0.30 & \hspace{7mm} -0&13 &  0.73 &  1.68 &  0.63 \\[0.4mm]
       & 1s $\rightarrow$ 2$e$/3p   & \hspace{4mm} 402&12$^{d}$   &  0.23 & \hspace{7mm} -0&04 &  0.72 &  1.66 &  0.47 \\[0.4mm]
       & 1s $\rightarrow$ 5$a_1$/3p & \hspace{4mm} 402&65$^{d}$   &  0.19 & \hspace{7mm}  0&23 &  0.85 &   -   &  1.00 \\[0.4mm]
OH$_2$ & 1s $\rightarrow$ 4$a_1$/3s & \hspace{4mm} 533&62$^{e}$   &  0.21 & \hspace{7mm}  0&0  &  0.78 &  2.06 &  0.53 \\[0.4mm]
       & 1s $\rightarrow$ 2$b_2$/3p & \hspace{4mm} 535&51$^{e}$   &  0.31 & \hspace{7mm} -0&1  &  0.70 &  1.96 &     - \\[1.0mm]
\multicolumn{10}{l}{ \textit{transition energies from the lowest core excited state}:\rule{0pt}{3.4mm} } \\[2mm]
CH$_4$ & 3$a_1$/3s $\rightarrow$ 2$t_2$/3p &         1&30     & -0.10 & \hspace{7mm}  0&2  &   -   &  -    &     - \\[0.4mm]
NH$_3$ & 4$a_1$/3s $\rightarrow$ 2$e$/3p   &         1&67     & -0.07 & \hspace{7mm}  0&09 & -0.01 & -0.02 & -0.16 \\[0.4mm]
       & 4$a_1$/3s $\rightarrow$ 5$a_1$/3p &         2&20     & -0.11 & \hspace{7mm}  0&36 &  0.12 &  -    &  0.37 \\[0.4mm]
OH$_2$ & 4$a_1$/3s $\rightarrow$ 2$b_2$/3p &         1&89     &  0.10 & \hspace{7mm} -0&1  & -0.08 & -0.10 &     - \\[1.6mm]
    \hline
    \multicolumn{10}{l}{${}^{a}$Adjustments for relativistic effects \cite{RelativisticCorrections}
                                of 0.10, 0.21, and 0.38 eV were applied for C, N, and O, respectively.\rule{0pt}{3.2mm}} \\
    \multicolumn{10}{l}{${}^{b}$Adjusted for one $\nu_4$ vibrational quanta.\cite{10e:EXP}\rule{0pt}{3.2mm}
        \hspace{6mm}  ${}^{d}$Taken from Table 3 of Schirmer et al.\cite{10e:EXP}\rule{0pt}{3.2mm}} \\
    \multicolumn{10}{l}{${}^{c}$Taken from Table 6 of Schirmer et al.\cite{10e:EXP}\rule{0pt}{3.2mm}
        \hspace{8.7mm}  ${}^{e}$Taken from Figure 1 of Schirmer et al.\cite{10e:EXP}\rule{0pt}{3.2mm}}
\end{tabular}
\end{table*}

In practice, we take the following approach to achieve energy
differences between variance-matched wave functions.
First, in the ground state, we discard all determinants
whose ORMAS CI coefficients have absolute values below 0.02
when constructing our VMC ansatz, which leads to
ground state wave functions with fewer than ten determinants
in each of the three molecules considered here.
After the ground state VMC optimization (which uses the same
target function as above), we then construct and optimize
an initial excited state ansatz.
Based on its variance relative to the ground state, we then
choose two additional expansion lengths for the excited state
and optimize them.
The idea is to straddle the ground state variance to allow
for simple interpolation,
as shown in Figure \ref{fig:vm_methane}.
Note that this linear regression approach is slightly
different than the nonlinear fitting used previously.
\cite{Robinson:2017}
We prefer the approach of Figure \ref{fig:vm_methane}
both for its simplicity and because it removes
the somewhat arbitrary choice of nonlinear fitting function.
So long as one is interpolating over a small distance,
a linear approximation is both straightforward and reasonable.
In Section \ref{sec::check_vm} below, we explicitly test
how sensitive this approach is to increasing the interpolation
distance and find that in practice the sensitivity is quite low.

\section{Results}
\label{sec:results}

\subsection{Transition Energies}
\label{sec::energies}

As a preliminary test of this VMC approach to core excitations,
we have applied it to the heavily studied ten electron series
of methane, ammonia, and water.
As seen in Table \ref{tab::energies}, the approach consistently
predicts ground-to-excited transition energies that are 0.2
to 0.3 eV above experiment once relativistic effects
\cite{RelativisticCorrections} are accounted for.
We suspect that this tendency to error slightly high may be
due to the fact that our ORMAS excited state calculations
are being done in an unrelaxed orbital basis and so likely
do a worse job than the corresponding ground state
calculations at predicting which determinants will be
most important in the final VMC orbital basis.
Although unrelaxed orbitals were used intentionally in order
to provide a strenuous test of our VMC optimization, it will
clearly be worthwhile in future to test the efficacy of
performing ORMAS CI for excited states in a relaxed orbital basis.
Although VMC does bias slightly high, the consistency with which
it does so is interesting in light of the
fact that the VMC optimization gets harder as the nuclei
get heavier, as one would expect.
For example, the per-electron variance at the end of our
H$_2$O optimizations is roughly twice the per-electron variance
achieved in CH$_4$, suggesting that in H$_2$O our wave functions
are less accurate.
Variance matching appears to do its job and make up for this,
as the accuracies of the predicted ground-to-excited and
excited-to-excited transition energies in all three molecules
are very similar.

Comparing with other recent theoretical approaches to core
excitations, we find the accuracy of VMC to be highly competitive.
For ground-to-excited transition energies, VMC substantially
outperforms equation of motion coupled cluster and NOCIS.
When using the SCAN functional, ROKS is if anything even
more accurate than VMC for these ground-to-excited
transitions, although its error is less systematic.
Other density functionals have been reported to be less
accurate than SCAN for core excitations, \cite{Dip:2020:roks_core}
although how much varies significantly across functionals
of different types.
For excited-to-excited transition energies, which when looking
at spectra are what determine peak separations, VMC's accuracy
is similar to equation of motion coupled cluster and
better than ROKS and NOCIS.
Thus, in these molecules, we find that VMC appears quite capable
of fulfilling its traditional role as a theoretical benchmark,
being competitive with both the best available method for
ground-to-excited transitions and the best available methods
for excited-to-excited transitions.
When thinking about a method's possible future value as a
benchmark method in systems without clear experimental data,
it is important to consider how systematically improvable
a method is in addition to how accurate it is.
Unlike the other methods in Table \ref{tab::energies},
this type of VMC approach is quite straightforward to
improve systematically by simply enlarging the
determinant expansion. \cite{scemama2016}

\begin{figure*}
    \centering
    \includegraphics[width=0.92\textwidth]{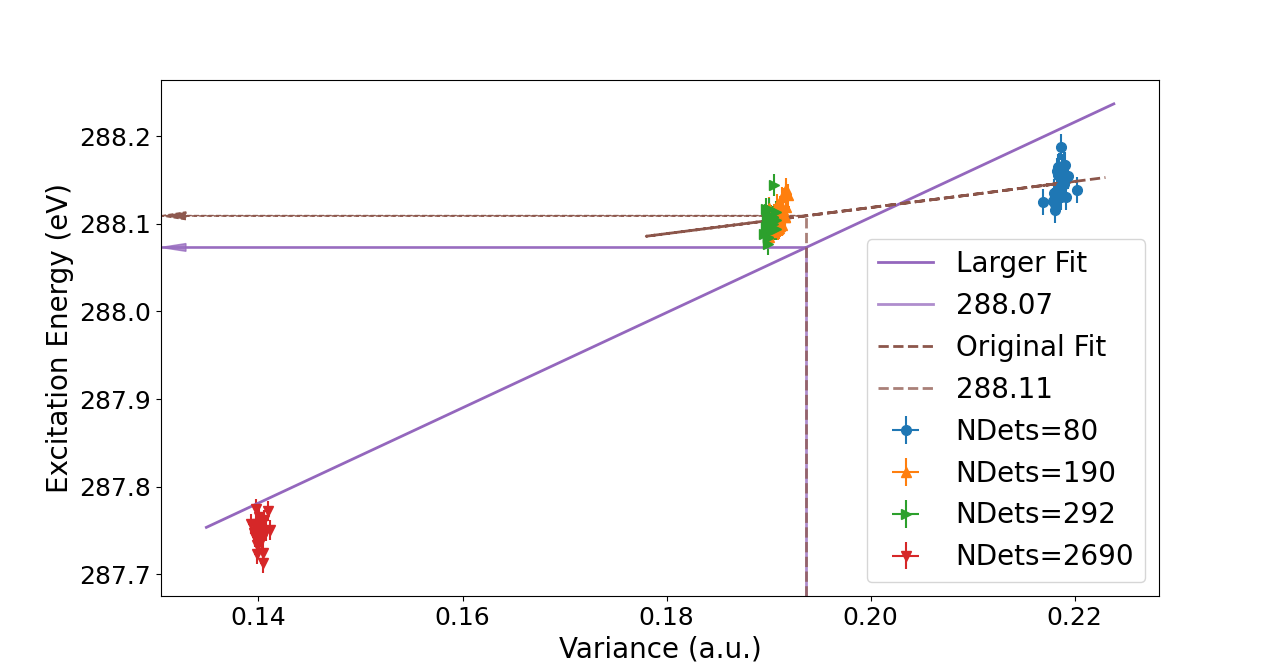}
    \caption{A comparison of difference variance matching linear regressions for the 2$t_2/3p$ excitation in methane.  The original fit is as reported in Table \ref{tab::energies}, while the larger fit includes an additional set of points from a significantly larger determinant expansion.  Even with two different fits, the excitation energy prediction changes by less than 0.05 eV.
    A third fit (not shown) that includes only the three larger wave
    functions gives a prediction within 0.02 eV of the original.
    \label{fig:xl}
    }
\end{figure*}


\subsection{Robustness of Variance Matching}
\label{sec::check_vm}

While the accuracies seen in the previous section suggest
our variance matching procedure is working well, we
nonetheless feel it is important to test its
sensitivity to the degree of interpolation employed.
It is difficult to predict a priori what excited state expansion
lengths will nicely straddle the ground state variance, and so
unless a large number of excited state optimizations are done,
how tightly the excited states straddle the ground state
will not be controlled systematically.
To test how sensitive variance matching is to the degree
of interpolation, we have therefore performed an additional
methane optimization with a much larger determinant expansion
and compared interpolations with and without it in order to
see the effect of interpolating over a wider variance range.

As seen in Figure \ref{fig:xl}, the excitation energy prediction
changes by less than 0.05 eV when including this extra excited
state calculation in the linear regression.
Notably, the variance range is now large enough that a clear
nonlinearity can be seen in the relationship between energy
and variance.
Nonetheless, a simple linear regression still gives almost the
same answer as before.
If one wanted to improve the suitability of using a linear
regression, the least accurate excited state wave function
could be omitted, which would lead to a linear fit over a
somewhat smaller variance range.
As the energy vs variance should be a smooth function,
linear fits should be increasingly appropriate as the
variance range is reduced.
This approach leads to an excitation energy prediction
that is even closer (now within 0.02 eV)
to the original interpolation
(i.e.\ the one without the 2690-determinant expansion).
Thus, different interpolations make little difference here,
although we do see that predictions are slightly more
consistent when the linear regression is done over
shorter ranges.

\section{Conclusion}
\label{sec:conclusion}

We have presented a systematically-improvable approach to core excitation energies
that accounts for correlation and orbital relaxation while explicitly balancing
the accuracies of the ground and excited state wave functions.
The approach involves a careful choice of basis set and electron-nuclear correlation
factor, a straightforward restricted active space approach for generating an initial
determinant expansion, variational Monte Carlo, and the use of the variance matching
technique for enhancing error cancellation.
As the computational bottleneck is clearly the Monte Carlo optimization,
this is by no means a low-cost approach, but
the very high accuracies it displays in our preliminary tests on water, ammonia,
and methane suggest that it should be useful for benchmarking other theoretical
methods in systems where experimental data is absent or less reliable.
Notably, other recently-developed theoretical methods
(ROKS, fc-CVS-EOM-CCSD, and CVS-LR-CCSD)
that offer comparable accuracies for ground-to-excited and/or excited-to-excited
transition energies are much harder to improve systematically.


Looking forward, many extensions to this approach present themselves.
For starters, it seems obvious that in future, an orbital relaxed basis should be
used for the excited state restricted active space calculations, as this will
almost certainly improve their ability to predict which determinants will ultimately
be important for correlation recovery.
Another straightforward step would be to use our approach to prepare nodal surfaces
for diffusion Monte Carlo, although some caution is in order here as this could in
principle at least spoil the error cancellation that variance matching provides.
Of course, extending variance matching itself to projector Monte Carlo methods may help.
In terms of potentially useful benchmarking applications, the area of doublet radical
core states has received increasing theoretical attention lately
\cite{roemelt2013,wenzel2014,Coriani:2015,oosterbaan2019,Dip:2020:radical_core}
and is an area where experimental data is less commonly available.
Another promising application area is in pump-probe experiments aimed at
photochemical processes (e.g.\ in DNA bases\cite{wolf2017}),
where a molecule with an existing valence excitation is subjected to an additional
core excitation.
The resulting doubly excited states are especially challenging for modern quantum
chemistry, but do not present any formal problems
for the Monte Carlo approach developed here.
In cases like this where the cost of Monte Carlo is not prohibitive,
it would be interesting to employ it to help benchmark more affordable
theories in this challenging area.




$\vspace{1mm}$

\noindent
\textit{Acknowledgements} ---
This work was supported by the
Office of Science, Office of Basic Energy Sciences,
the U.S.\ Department of Energy,
Contract No.\ {DE-AC02-05CH11231}.
Calculations used
the LBNL Lawrencium computing cluster.

$\vspace{1mm}$

\noindent
\textit{Data Availability Statement} ---
The data that supports the findings of this study
are available within the article.

\bibliography{main}

\end{document}